\documentclass[conference]{IEEEtran}
\usepackage{graphicx,psfrag,epsfig,epsf,latexsym,hhline,amsmath,amssymb,multirow}
\usepackage[usenames,dvipsnames]{pstricks}
\usepackage{pst-plot}
\usepackage{caption}
\usepackage{subcaption}
\usepackage{enumitem} 
\usepackage{amsmath} 
\usepackage{graphicx}
\usepackage{color}
\usepackage{hyperref}
\usepackage{graphicx}
\usepackage{amsthm}
\usepackage{algorithm}
\usepackage{algpseudocode}
\usepackage[noadjust]{cite}
\usepackage{blindtext}
\usepackage{etoolbox,tcolorbox}
\usepackage{stfloats}
\usepackage{tikz}
\usetikzlibrary{chains,shapes.multipart}
\usetikzlibrary{shapes,calc,fit}
\usetikzlibrary{automata,positioning}
\graphicspath{ {figures/} }

\newcounter{cntr}
\tikzset{
    queuei/.pic={
  \stepcounter{cntr}
        \node[outer sep=0pt,draw,rectangle split,rectangle split horizontal,minimum height=10pt,rectangle split parts=3] (queue-\thecntr) [pic actions] {};
        \draw
          (queue-\thecntr.north west) -- ++(-10pt,0)
          (queue-\thecntr.south west) -- ++(-10pt,0);
    },
}

\input{Jerry.def}

\begin{document}
\title{Outage Analysis of Age-of-Information for Multi-Source Systems}
\author{Guan-Yu Lin$^\dag$, Yu-Chih Huang$^\dag$, and Yu-Pin Hsu$^*$\\
$^\dag$Institute of Communications Engineering, National Yang Ming Chiao Tung University, Hsinchu, Taiwan\\
$^*$Department of Communication Engineering, National Taipei University, New Taipei City, Taiwan\\
E-mail: \{casperlin.ee09@nycu.edu.tw, jerryhuang@nycu.edu.tw, yupinhsu@mail.ntpu.edu.tw\} }

\maketitle

\begin{abstract}
Age of information (AoI) is an effective performance metric measuring the freshness of information and is popular for applications involving status update. Most of the existing works have adopted average AoI as the metric, which cannot provide strict performance guarantees. In this work, the outage probability of the peak AoI exceeding a given threshold is analyzed in a multi-source system under round robin scheduling. Two queueing disciplines are considered, namely the first-come-first-serve (FCFS) queue and the single packet queue. For FCFS, upper and lower bounds on the outage probability are derived which coincides asymptotically, characterizing its true scaling. For the single packet queue, an upper bound is derived whose effectiveness is validated by the simulation results. The analysis concretizes the common belief that single packet queueing has a better AoI performance than FCFS. Moreover, it also reveals that the two disciplines would have similar asymptotic performance when the inter-arrival time is much larger than the total transmission time.

\end{abstract}





\section{Introduction}\label{sec:intro}
In recent years, the success of many new applications in the Internet of Things (IoT) has hinged greatly on the timeliness of information. Accordingly, the age of information (AoI) was first proposed in \cite{Kaul12_age_provide} as a new performance metric to measure the freshness of information. AoI measures the amount of time elapsed since the generation of the latest updated packet until the present and is fundamentally different from the notion of delay. It has also been demonstrated in \cite{Yates21} that delay optimal network design does not minimize AoI. Therefore, designs and analyses for delay do not necessarily carry over to those for AoI straightforwardly. 
Since \cite{Kaul12_age_provide}, there has been a lot of work using AoI as the performance metric. The reader is referred to \cite{Yates21} for a comprehensive survey.


Another important feature of IoT is that there are typically a number of heterogeneous applications/devices sharing the system. For such a scenario, scheduling becomes inevitable. With average AoI as the metric, scheduling has been investigated in \cite{Kadota18,Maatouk21,Jiang19,Yang21,Hsu20}. In \cite{Kadota18}, three low-complexity scheduling policies were considered and analyzed, namely the randomized policy, the Max-Weight policy and the Whittle's index policy. In \cite{Maatouk21},  the Whittle's index policy was further proved to minimize the average AoI in the multi-source regime. In \cite{Jiang19}, the round robin policy was investigated under single packet queue (i.e., the queue length is one and it preempts the packets whenever there is a new packet arrived) and was shown to be optimal among all arrival-independent policies. In \cite{Yang21}, a distributed algorithm for a Poisson bipolar network was proposed and analyzed. In \cite{Hsu20}, the authors developed a structural MDP scheduling algorithm and an index scheduling algorithm to minimize the average AoI.


Despite many successes in the AoI literature thus far, most of them focus solely on using the average AoI as the metric. Although minimizing average AoI more or less implies reducing the probability of violating AoI threshold, it cannot be directly converted to performance guarantee. Therefore, these results are not directly applicable to applications that have strict age requirements. To cope with this, in this work, we aim at analyzing the distribution of AoI in a multi-source system. Specifically, for any given threshold, we analyze the outage probability that the peak AoI is greater than the described threshold. Due to its simplicity and effectiveness demonstrated in \cite{Jiang19}, the round-robin policy is considered. Moreover, two queueing disciplines, namely the first come first serve (FCFS) queue and the single packet queue, are analyzed. 

Note that AoI in various queueing systems has been studied before. For example, \cite{Kaul12_LCFS} considered the last come first serve (LCFS) queue, \cite{Soysal21} considered the G/G/1 queue with the FCFS and LCFS, and \cite{Kam16} analyzed average AoI under M/M/1, M/M/2, and M/M/$\infty$ system. However, these works considered the single source system and focused solely on the average AoI. It must be emphasized that there has been a series of works \cite{Jaya21_min_achieve,Jaya21_multi_hop,Limei21,Seo19} analyzing the distribution of peak AoI. However, these works again focused solely on the single source system and the analysis therein are not directly applicable to the multi-source scenario. Given the existing literature, to the best of our knowledge, the present work is the first that attempts to understand the distribution of AoI in the multi-source system. 

As the first attempt, in this work, we consider that the sources' packets arrive in bulk periodically. This assumption is not merely for easing the analysis, but also to address a practical scenario where resource is pre-allocated in a periodic fashion (see semi-persistent scheduling in NB-IoT systems \cite{Sergey19,Wang17} for example.) On the other hand, the transmission time is stochastic and can have any distribution as long as the first two moments exist. We successfully analyze the outage probability of the peak AoI exceeding a predefined threshold for a multi-source system under round robin scheduling. For FCFS, an upper bound and a lower bound on the outage probability for any given threshold are derived. These bounds turn out to coincide asymptotically when the number of sources tend to infinity, characterizing the true scaling for this case. For the single packet queue case, an upper bound on the outage probability is derived. Although a tight lower bound is lacking, our simulation results indicate the same scaling with the derived upper bound when the number of sources is large. With our analysis, two conclusions can be drawn: 1) the common belief that the single packet queue has a better AoI performance than FCFS is made precise even in terms of the outage probability; and 2) the two queueing disciplines result in the same asymptotic scaling when the inter-arrival time is much larger than the total transmission time.

The rest of the paper is organized as follows. In Section \ref{sec:system model}, we illustrate the network model, the age model, and the problem we study. In Section \ref{sec:FCFS}, we provide the analysis of the outage probability for FCFS queue. 
Our analysis for the single packet queue is then provided in Section \ref{sec:LCFS}.
In Section \ref{sec:numerical and simulation}, we validate our analysis with simulations and compare the difference between two queueing disciplines. Finally, in Section \ref{sec:conclusion}, we conclude the paper. 

\subsection{Notation}
Throughout the paper, constants and random variables are written in lowercase and uppercase, respectively, for example, $x$ and $X$. For a real constant $x$, we use $(x)^+$ to denote $\max\{0,x\}$. Also, for a positive integer $n$, $[n]$ is an abbreviation of the set $\{1, 2, \ldots, n\}$.



\section{System Model and Problem Formulation} \label{sec:system model}
In this section, we describe the network model in Section~\ref{subsec:net_model}, followed by the notion of AoI and a description of the problem we attempt to solve in Section~\ref{subsec:AoI}.

\subsection{Network Model}\label{subsec:net_model}
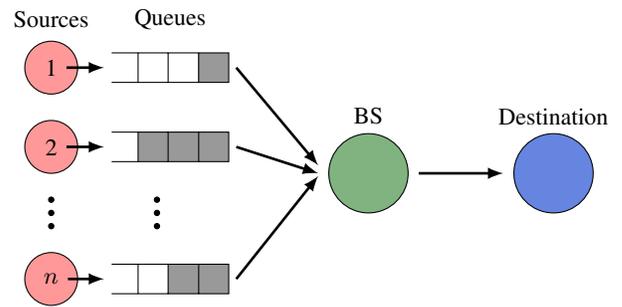
\begin{figure}
\centering
\begin{tikzpicture}[>=latex]
\def\pos{0}
\draw [fill=red!40] (\pos,100pt) circle [radius=10pt] node (s1) {\small 1};
\draw [fill=red!40] (\pos,70pt) circle [radius=10pt] node (s2) {\small 2};
\draw [fill=red!40] (\pos,20pt) circle [radius=10pt] node (sn) {\small $n$};
\foreach \idx in {10, 15, 20}
    \filldraw [black] ([yshift=10+\idx pt]sn) circle (1pt);

\path
(\pos+50pt,100pt) pic[rectangle split part fill={white,white,gray!80}] {queuei=1}
(\pos+50pt,70pt) pic[rectangle split part fill={gray!80,gray!80,gray!80}] {queuei=2}
(\pos+50pt,20pt) pic[rectangle split part fill={white,gray!80,gray!80}] {queuei=n};
\path
(\pos+40pt,100pt) coordinate (Q1)
(\pos+40pt,70pt) coordinate (Q2)
(\pos+40pt,20pt) coordinate (Qn);
\foreach \idx in {10, 15, 20}
    \filldraw [black] ([yshift=10+\idx pt]Qn) circle (1pt);

\draw[line width=1pt, ->] ([xshift=10pt]s1)--([xshift=-20pt]Q1);
\draw[line width=1pt, ->] ([xshift=10pt]s2)--([xshift=-20pt]Q2);
\draw[line width=1pt, ->] ([xshift=10pt]sn)--([xshift=-20pt]Qn);

\draw [fill=black!60!green!50]  (\pos+120pt,60pt) circle [radius=15pt] node{} coordinate (BS);

\draw[line width=1pt, ->] ([xshift=30pt]Q1)--([xshift=-19pt, yshift=3pt]BS);
\draw[line width=1pt, ->] ([xshift=30pt]Q2)--([xshift=-19pt, yshift=0pt]BS);
\draw[line width=1pt, ->] ([xshift=30pt]Qn)--([xshift=-19pt, yshift=-1pt]BS);



\draw [fill=green!20!blue!60] (\pos+190pt,60pt) circle [radius=15pt] node{} coordinate (Dn);

\draw[line width=1pt, ->] ([xshift=19pt]BS)--([xshift=-19]Dn);

\begin{scope}
\node [anchor=south] at ([yshift=5pt]s1.north) {\small Sources};
\node [anchor=south] at ([xshift=5pt, yshift=11pt]Q1.north) {\small Queues};
\node [anchor=south] at ([yshift=15pt]BS.north) {\small BS};
\node [anchor=south] at ([yshift=15pt]Dn.north) {\small Destination};
\end{scope}
\end{tikzpicture}
\caption{An illustration of the network model.}\label{fig:illustrate network}
\end{figure}

We consider an information update system shown in Fig.~\ref{fig:illustrate network}, where $n$ sources update the status of information to a destination through a base station (BS).\footnote{The model is equivalent to having $n$ destinations where each source is requested by a destination.} 
The BS samples every source every $nb$ time units, where $b$ is a constant. That is, the $n$ sources arrive in bulk in a deterministic and periodic fashion. Moreover, we denote by $S_i(k)$ the arrival time of the $k$-th updated packet of source $i$.

The BS maintains a queue for each source, where (some of the) unserved packets of source $i\in[n]$ are stored in the $i$-th queue. In this paper, two service disciplines are considered, namely the FCFS queue and the single packet queue. In FCFS, the queue size is infinite and the BS serves the earliest arriving packets that have not been served. For single packet queue, to prevent BS from sending stale information, the last arriving packet is served and queue size is one; moreover, the previous packet in the queue is preempted. 

We denote by $V_i(k)$ the transmission time of the $k$-th scheduled packet (also called the $k$-th round of transmission) of the source $i\in[n]$. $V_i(k)$ is assumed to be stochastic and is independent and identically distributed (i.i.d) according to a specific distribution whose first two moments exist. 
In addition to the queueing discipline that decides how the packets of a user should be served, there is a scheduling policy $\pi$ which governs the order of service among users.



\subsection{Age of Information and Problem Formulation}\label{subsec:AoI}
We now describe the definition of AoI. Under scheduling policy $\pi$, let the departure time of the $k$-th scheduled packet of the source $i$ be $D^\pi_{i}(k)$ and let $U^\pi_i(t) = \{ S_i(k)| k = \arg \max_k D^\pi_i(k)\leq t \}$ be the generation time of the latest packet transmitted to the destination by time $t$. Then, AoI for source $i$ at time $t$ is defined by $\Delta^\pi_i(t) = t-U^\pi_i(t)$ and is depicted in Fig.~\ref{fig:age evolution}. 
The peak age of the $k$-th updated packet of user $i$ is defined as
\begin{equation}\label{eqn:A(k)}
A^\pi_i(k) = D^\pi_i(k)-S_i(k-1),
\end{equation}
which captures the time from the generation of the $(k-1)$-th scheduled packet of the source $i$ to the successful departure of the $k$-th scheduled packet from the same source.

Note that we use the superscript $\pi$ to emphasize that those random variables depend on the underlying scheduling policy. In this paper, we focus solely on the round-robin scheduling policy defined below and we drop the superscript $\pi$ from this point onward.

\begin{define}[Round-robin policy]
Let $\tau_i(t)$ be the time elapsed since the last time the source $i$ was updated to the current time $t$. Then, the round-robin policy serves the packet in the queue of the source $j = \arg \max_{i\in [n]} \tau_i(t)$. If there is no packet in that queue, the BS waits until the next arrival to the queue.
\end{define}

Unlike most of the work in the AoI literature considering the long-term average of the peak AoI, to have a better characterization of the behavior of the peak AoI, we aim to analyze the outage probability defined as follows. 
\begin{define}[Outage probability]
    Note that $Pr(A_i(k)>nx)$ is the outage probability that the peak AoI of the $k$-th scheduled packet from source $i$ is above a given threshold $nx$. For $i\in[n]$, we define the outage probability of the source $i$ as
    \begin{equation}
        P_{out,i}^n(x) \triangleq \lim_{k\to\infty}Pr(A_i(k) \geq nx).
    \end{equation}
\end{define}
Having defined the outage probability, we then define the asymptotic decay rate, which provides a guarantee on the decay rate of the outage probability in a large-user system.
\begin{define}[Asymptotic decay rate]
The asymptotic decay rate $\gamma(x)$ under the threshold $x$ is said to be the rate at which the peak age decays as the total number of sources in the system increases to infinite. That is,
\begin{equation}\label{eqn:age requirement}
    \lim_{n \to \infty} \frac{1}{n} P_{out,i}^n(x) = \gamma(x).
\end{equation}
\end{define}

In this work, we also aim to analyze the asymptotic outage probability characterization for the round-robin scheduling policy when the number $n$ of sources is large.
\section{FCFS queue}\label{sec:FCFS}

The FCFS queue system is considered in this section. We first provide the age evolution together with some insights and explanations in Section \ref{subsec:age analysis FCFS}. Based on the age evolution formula, we then derive an upper bound and a lower bound on the outage probability of peak age, which allows us to characterize the asymptotic decay rate for the large-system scenario in Section \ref{subsec:asymptotic analysis FCFS}.

\subsection{Age analysis under FCFS queue}\label{subsec:age analysis FCFS}
\begin{figure}
	\center{\includegraphics[width=0.4\textwidth]
    {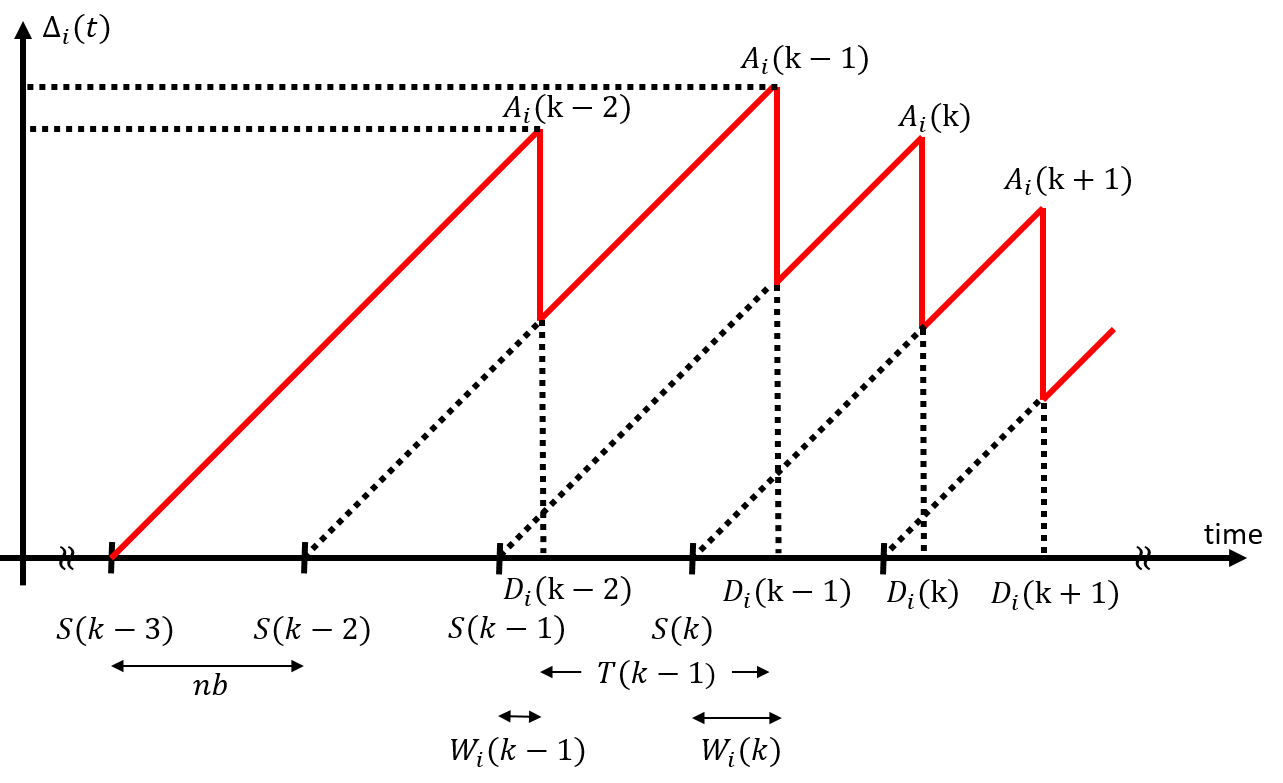}}
   	\caption{ Age evolution in a multi-source system.}
   	\label{fig:age evolution}
   	\vspace{-15pt}
\end{figure}

In FCFS, since every packet is stored in the queue until it gets served and the round robin is employed, the BS updates the packets in the same batch for all $i\in[n]$ in the $k$-th round. i.e., $S_i(k)=S(k)=k\cdot nb$. Moreover, $D_i(k)$ is equal to the sum of the arrival time of the $k$-th packet, the time this packet spends in the queue, and the transmission time of this packet. i.e., 
\begin{equation}\label{eqn:D_i_FCFS}
    D_i(k) = S(k) + W_i(k) + V_i(k),
\end{equation}
where $W_i(k)$ is the waiting time at the queue of the $k$-th packet from the source $i$. An illustration of the relationship among these random variables can be found in Fig. \ref{fig:age evolution}. Moreover, under the round-robin policy, the waiting time can be expressed by
\begin{equation}\label{eqn:W_i_FCFS}
    W_i(k) = W_{i-1}(k)+V_{i-1}(k)=W_1(k)+\sum_{u=1}^{i-1}V_u(k).
\end{equation}
Plugging \eqref{eqn:D_i_FCFS} and \eqref{eqn:W_i_FCFS} into \eqref{eqn:A(k)} yields
\begin{align}\label{eqn:A_i_FCFS}
	A_i(k) = W_1(k) + \sum_{u=1}^i V_u(k) + nb.
	\end{align}


The next proposition further expresses $W_1(k)$ in a recursive form so that we can get the formulation of peak age in Lemma~\ref{lemma:peak_age_formulation_FCFS}.
\begin{proposition}\label{prop:waiting_time_FCFS}
	The waiting time of the $k$-th updated packet from the first source can be expressed recursively as follows,
		\begin{equation}\label{eqn:W(k)}
			W_1(k) = \left(W_1(k-1) + \sum_{u=1}^n V_u(k-1) - nb\right)^+.
		\end{equation}
\end{proposition}
\begin{IEEEproof}
	Since the $k$-th packet from source 1 will be served after {\it all} $(k-1)$-th arriving packets are served, $W_1(k)$ depends on $D_n(k-1)$. If $D_n(k-1) > S(k)$, the BS will serve the $k$-th packet from source 1 immediately after the successful departure of the $k-1$-th packet from source $n$. We have
    \begin{align}\label{eqn:express W(k) by W(k-1)}
		W_1(k) &= D_n(k-1) - S(k) \nonumber\\
		&\overset{(a)}{=} S(k-1) + W_n(k-1) + V_n(k-1) - S(k) \nonumber\\
		&= W_n(k-1) + V_n(k-1) - nb \nonumber\\
		&\overset{(b)}{=} W_1(k-1) + \sum_{u=1}^{n} V_u(k-1) - nb,
    	\end{align}
    where (a) follows from \eqref{eqn:D_i_FCFS} and (b) follows from \eqref{eqn:W_i_FCFS}. If $D_n(k-1)\leq S(k)$, the BS is idle after transmitting the $(k-1)$-th packet of source $n$ and we can immediately transmit the $k$-th packet at the packet's arrival. Therefore, the waiting time is zero, which is captured again by \eqref{eqn:W(k)} as $W_1(k-1) + \sum_{u=1}^{n} V_u(k-1) - nb=D_n(k-1)-S(k)\leq 0$.  
\end{IEEEproof}
By plugging \eqref{eqn:W(k)} into \ref{eqn:A_i_FCFS} and some manipulation, we arrive at the evolution of AoI in the following lemma.

\begin{lemma}\label{lemma:peak_age_formulation_FCFS}
The peak age of the $k$-th updated packet of source $i$ under FCFS is
		\begin{align}\label{eqn:peak_age_formulation_FCFS}
			A_i(k) = \max_{1\leq s \leq k} &\left\{ \sum_{r=s}^{k-1} \sum_{u=1}^n V_u(r) + \sum_{u=1}^i V_u(k) \right. \nonumber\\
			&\hspace{30pt}\left. \vphantom{\sum_{u=1}^i V_u(k)} - (k-s-1)nb \right\}.
		\end{align}
\end{lemma}
\begin{IEEEproof}
    This lemma can be proved by recursively plugging \eqref{eqn:W(k)} into \eqref{eqn:A_i_FCFS}.
    The full proof can be found in Appendix~\ref{apx:proof_peak_age_formulation_FCFS}. 
\end{IEEEproof}

Intuitively, the AoI evolution above can be reasoned as follows. For a particular $1\leq s\leq k$ such that every round has $\sum_{u=1}^nV_u(r)\geq nb$, the waiting time accumulated since round $s$ can be expressed as the sum of $\sum_{r=s}^{k-1}\sum_{u=1}^n V_u(r)$ the transmission time of rounds $s$ to $k-1$ and $\sum_{u=1}^{i-1} V_u(k)$ that of users 1 to $i-1$ in the round $k$ minus $(k-s)nb$. The resulting AoI lower bound (it is a lower bound as we only accumulated the waiting time since round $s$) is then the sum of this waiting time, the transmission time $V_i(k)$, and $nb$. This lower bound becomes exact when we take the maximization over $1\leq s\leq k$.

\subsection{outage analysis under FCFS queue}\label{subsec:asymptotic analysis FCFS}
We now provide an upper and a lower bounds on the outage probability.
Let $\Lambda_v(\theta) = \log \mathbb{E} \left[e^{\theta V_i(k)}\right]$ be the log-moment generating function of random variable $V_i(k)$. We first present the upper bound in the following theorem.
\begin{theorem}\label{thm:upper_bound_FCFS}
The outage probability of the $k$-th updated packet of source $i$ under FCFS queue is upper bounded as follows,
		\begin{equation}\label{eqn:upper_bound_FCFS}
			P_{out,i}^n(x) \leq c \cdot \exp \left[ -n \cdot \min_{r\geq 1}\ r \cdot I_v \left( \frac{x}{r}+\frac{r-2}{r}b\right) \right],
		\end{equation}
where $c>0$ is a constant and
\begin{equation}\label{eqn:I_v}
    I_v(x) = \sup_{\theta:\Lambda_v(\theta) - \theta b < 0} \left[ \theta x - \left( \frac{r+1-\alpha_i}{r} \right) \Lambda_v(\theta) \right],
\end{equation} 
with $\alpha_i = \frac{i}{n}$.
\end{theorem}
\begin{IEEEproof}
	To establish this theorem, we apply the union bound and the Chernoff bound to \eqref{eqn:peak_age_formulation_FCFS}. The complete proof is given in Appendix~\ref{apx:proof_upper_bound_FCFS}.
\end{IEEEproof}

The lower bound is presented in the following theorem.
\begin{theorem}\label{thm:lower_bound_FCFS}
For any $\epsilon>0$ and any $r>0$, the outage probability of the $k$-th updated packet of source $i$ under FCFS queue is lower bounded as follows,
		\begin{equation}\label{eqn:lower_bound_FCFS}
			P_{out,i}^n(x) \geq \exp \left[ -n \cdot r\cdot I_v \left( \frac{x}{r}+\frac{r-2}{r}b\right) + \epsilon \right],
		\end{equation}
where $I_v(x)$ is in \eqref{eqn:I_v}.
\end{theorem}
\begin{IEEEproof}
    This theorem can be proved by choosing an arbitrary $s$ in \eqref{eqn:peak_age_formulation_FCFS} and applying the Chernoff bound. The full proof can be found in Appendix~\ref{apx:proof_lower_bound_FCFS}.
\end{IEEEproof}

With the above two theorems, we can characterize the asymptotic scaling of the outage probability for any $x>0$ in the limit as $n$ tends to infinity.
\begin{corollary}\label{coro:asym_FCFS}
For any $x>0$, the asymptotic decay rate of source $i$ under FCFS queue is given by
		\begin{equation}\label{eqn:asym_FCFS}
			\lim_{n \to \infty} \frac{1}{n} \log P_{out,i}^n(x) = -\min_{r\geq 1} \  r\cdot I_v\left(\frac{x}{r}+\frac{r-2}{r}b\right).
		\end{equation}
\end{corollary}
These results will be further validated by simulations in Section \ref{sec:numerical and simulation}.

\section{Single packet queue}\label{sec:LCFS}
In this section, we turn our attention to the single packet queue case. We again analyze the evolution of AoI in Section~\ref{subsec:age analysis preemptive} and derive an upper bound on the outage probability in Section~\ref{subsec:asymptotic analysis preemptive}.


\subsection{Age analysis under single packet queue}\label{subsec:age analysis preemptive}
In the single packet queue case, when a packet arrives at the BS, it preempts the previous packet from the same source, i.e. the queue stores the latest packet. Intuitively, this prevents the BS from updating stale information, resulting in a smaller AoI as compared to FCFS. Our following analysis makes this intuition precise. 

Because of the preemption, the arrival time of the $k$-th updated packet for different sources can be different. The peak age of the $k$-th updated packet of source $i$ is
\begin{equation}\label{eqn:A(k) LCFS}
    A_i(k) = D_i(k) - S_i(k-1).
\end{equation}
Similar to the FCFS queue case, the departure time is given by
\begin{equation}\label{eqn:D_i_LCFS}
    D_i(k) = S_i(k) + W_i(k) + V_i(k).
\end{equation}
Plugging \eqref{eqn:D_i_LCFS} into \eqref{eqn:A(k) LCFS}, we have
\begin{equation}\label{eqn:A_i_LCFS}
    A_i(k) = W_i(k) + V_i(k) + S_i(k) - S_i(k-1).
\end{equation}

\begin{figure}
	\centering
    \begin{subfigure}[b]{3in}
	\includegraphics[width=\textwidth]{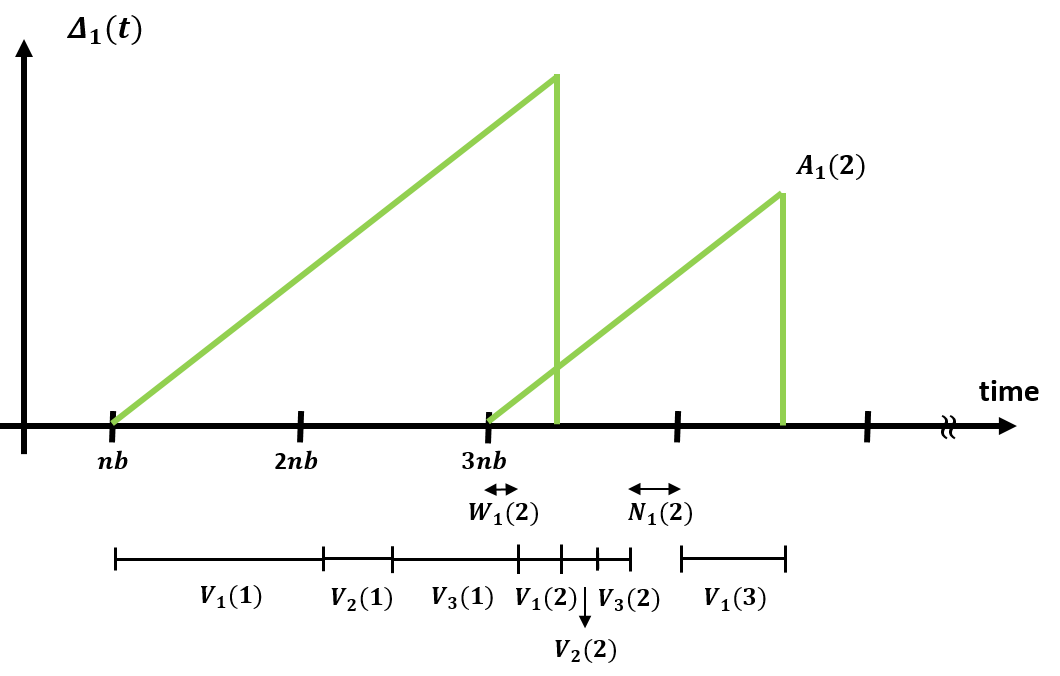}
   	\caption{\label{fig:LCFS_case1} Idle time $N_1(2)$ occurs after serving the third source.}
    \end{subfigure}
    \hfill
    \begin{subfigure}[b]{3in}
	\includegraphics[width=\textwidth]{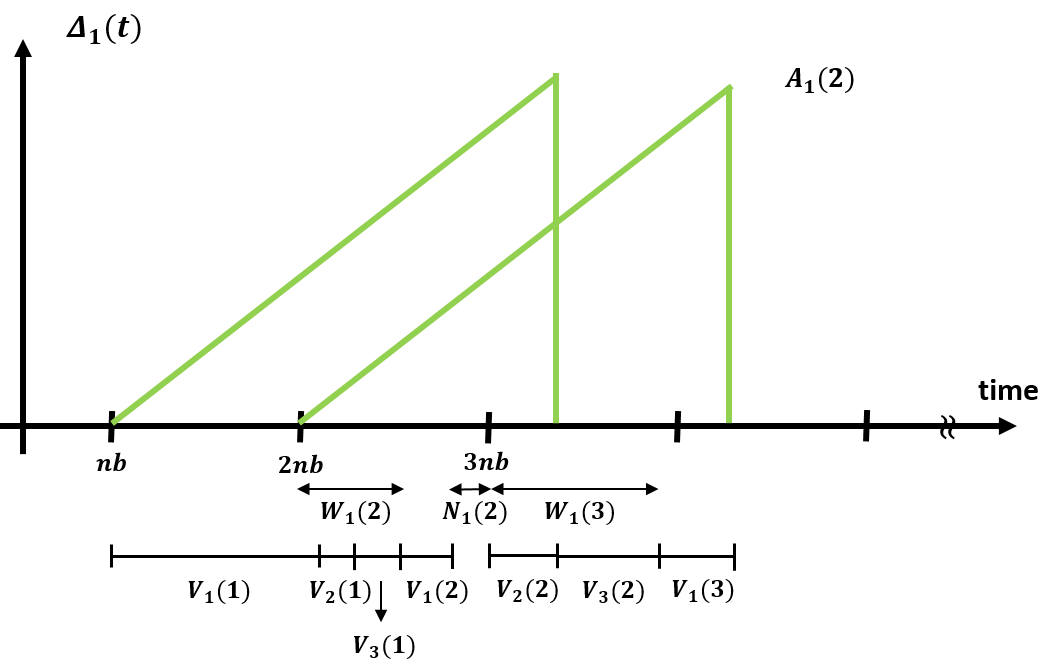}
   	\caption{\label{fig:LCFS_case2} Idle time $N_1(2)$ occurs after serving the first source.}
    \end{subfigure}
    \hfill
    \begin{subfigure}[b]{3in}
	\includegraphics[width=\textwidth]{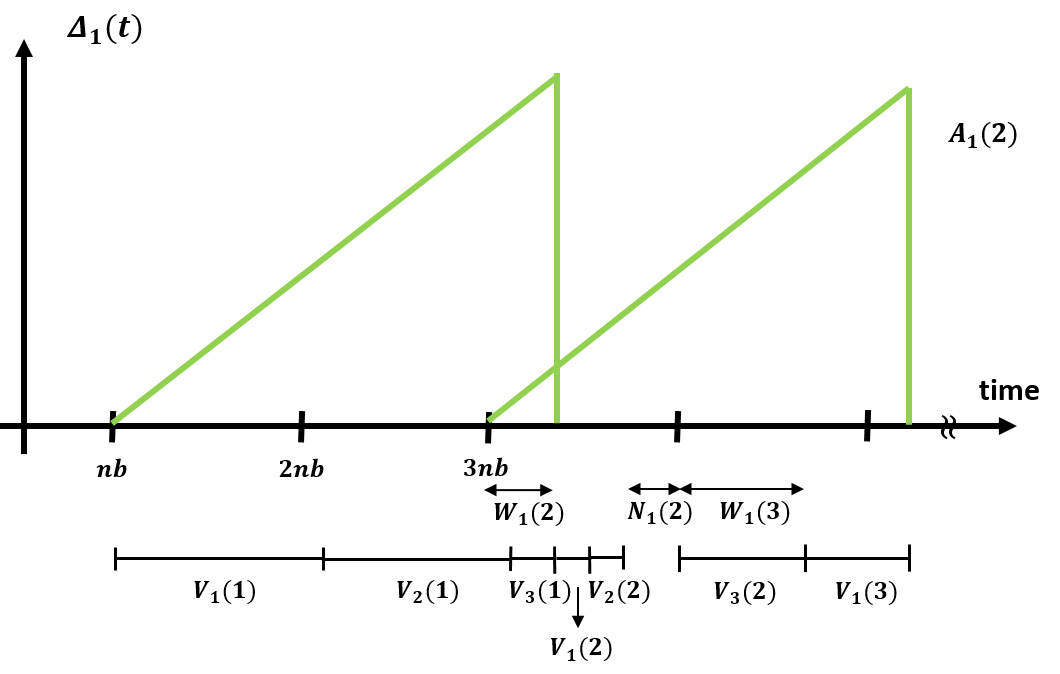}
   	\caption{\label{fig:LCFS_case3} Idle time $N_1(2)$ occurs after serving the second source.}
    \end{subfigure}
    \caption{An example of AoI in single packet queue.}
    \label{fig:LCFS_ex}
    \vspace{-20pt}
\end{figure}
    In the following lemma, we analyze the waiting time for the single packet queue case.\\
    \begin{lemma}\label{lemma:waiting_time_LCFS}
        The waiting time of the $k$-th updated packet for source $i$ can be expressed as follows.
        \begin{align}\label{eqn:W_i_LCFS}
            W_i(k) = & W_i(k-1) + \sum_{u=i}^n V_u(k-1) + N_i(k-1) \nonumber\\
            &+ \sum_{u=1}^{i-1} V_u(k) - \left( \vphantom{\sum_{u=i}^n V_u(k-1)} p_i(k-1)+1 \right) \cdot nb 
        \end{align}
        where $N_i(k-1)$ is the idle time occurring after serving the $(k-1)$-th update from source $i$ and $p_i(k-1) = \left\lfloor \frac{W_i(k-1) + \sum_{u=i}^n V_u(k-1) + N_i(k-1) + \sum_{u=1}^{i-1} V_u(k)-nb}{nb} \right\rfloor $ is the number of source $i$s' packets preempted between the $(k-1)$-th and the $k$-th updates.
    \end{lemma}
    \addtolength{\topmargin}{0.04in}
    \begin{IEEEproof}
        We first note that without preemption, the waiting time of the $k$-th updated packet from source $i$ would be
        \begin{equation}\label{eqn:W_i(k) by W_i(k-1)}
            W_i(k-1) + \sum_{u=i}^n V_u(k-1) + N_i(k-1) - nb + \sum_{u=1}^{i-1} V_u(k),
        \end{equation}
        where $W_i(k-1)$ is the waiting time in the previous round, $\sum_{u=i}^n V_u(k-1)$ is the transmission time for the packets in the previous round, $\sum_{u=1}^{i-1} V_u(k)$ is the transmission time of the $k$-th updated packets from the sources before $i$, and $N_i(k-1)$ is the idle time to fill the gap (if exists) between the finish of updating all round $k-1$ packets and the arrival of the next batch. Now, if $\eqref{eqn:W_i(k) by W_i(k-1)}\geq nb$, the $p_i(k-1)$ stale packets from this source will be preempted. This completes the proof.
    \end{IEEEproof}
    We use a three-source system in Figs.~\ref{fig:LCFS_ex} as an example to explain $N_i(k-1)$ and $p_i(k-1)$. In Fig.~\ref{fig:LCFS_case1}, the first round has a fairly large transmission time of $V_1(1)+V_2(1)+V_3(1)>2nb$; hence, $N_1(1)=0$ and $p_1(1)=1$. On the contrary, the second round has a short overall transmission time; hence, the BS has to wait $N_1(2)$ until the arrival of the next batch to start the service of source 1 in the third round. In this case, $p_1(2)=0$. In Fig.~\ref{fig:LCFS_case2}, we argue that the idle time $N_i(k-1)$ does not necessarily occur after serving an entire round. In this figure, we see that $V_1(1)$ is fairly large, which results in the preemption of one packet from the source 2 and 3. After $V_2(1)$, $V_3(1)$, and $V_1(2)$, the packets arrived at $2nb$ are all served and the BS again becomes idle. Fig.~\ref{fig:LCFS_case3} shows another case that the BS becomes idle after serving a packet from source 2. As shown in the above example, the behavior of the random variables $N_i(k-1)$ and $p_i(k-1)$ are difficult to track, making the analysis of the single packet queue quite challenging.
    
Next, we provide the formulation of the peak age of the $k$-th update of source $i$ in Lemma\ref{lemma:peak_age_formulation_preemption}.

\begin{lemma}\label{lemma:peak_age_formulation_preemption}
		The peak age of the $k$-th update of source $i$ under single packet queue can be expressed as follow,
		\begin{equation}\label{eqn:peak_age_preemptive}
		A_i(k) = W_i(k-1) + \sum_{u=i}^n V_u(k-1) + N_i(k-1) + \sum_{u=1}^i V_u(k).
		\end{equation}
\end{lemma}
\begin{IEEEproof}
    Plugging \eqref{eqn:W_i_LCFS} into \eqref{eqn:A_i_LCFS}, results in,
    \begin{align*}
        A_i(k) &= W_i(k-1) + \sum_{u=i}^n V_u(k-1) + N_i(k-1) + \sum_{u=1}^{i} V_u(k)\\
        &\quad - \left[ \vphantom{\sum_{u=i}^n V_u(k-1)} p_i(k-1)+1 \right] \cdot nb + S_i(k) - S_i(k-1)\\
        &\hspace{-0.12in}\overset{(a)}{=} W_i(k-1) + \sum_{u=i}^n V_u(k-1) + N_i(k-1) + \sum_{u=1}^{i} V_u(k).
    \end{align*}
    where, in (a), we know that $S_i(k) - S_i(k-1)$ is the duration time between the arrival time of the $k-1$th update and the $k$th update. It is equal to one inter-arrival time plus the number of preempted packets multiplied by the inter-arrival time which is the total packets we go through between updates.
\end{IEEEproof}

\subsection{Asymptotic scaling under single packet queue}\label{subsec:asymptotic analysis preemptive}
With the evolution of AoI above, we are now ready to derive an upper bound on the outage probability of peak age. 
  
\begin{theorem}\label{thm:upper_bound_preemptive}
The outage probability of source $i$ under single packet queue is upper bounded as follows,
		\begin{equation}\label{eqn:upper_bound_preemptive}
			P_{out,i}^n \leq \exp \left[ -n \cdot I_v^{(U)} \left( x-b \right) \right].
		\end{equation}
where 
\begin{equation}\label{eqn:rate function preemptive}
    I_v^{(U)}(x) = \sup_{\theta>0} \left[ \theta x - \left( \frac{n+1}{n} \right) \Lambda_v(\theta) \right].
\end{equation}
\end{theorem}
\begin{IEEEproof}
	To get an upper bound, we upper bound the waiting time in \eqref{eqn:A_i_LCFS} by the maximum value $nb$ which results in $N_i(k-1)=0$. We continue to use the Chernoff bound for obtaining the upper bound. Details of the proof are in Appendix~\ref{apx:proof_upper_bound_preemptive}.
\end{IEEEproof}

The above theorem provides an upper bound on the outage probability for any number of sources $n$ and any threshold $x$. Let the number of sources goes to infinity. The asymptotic probability is expressed in the following corollary.
\begin{corollary}\label{coro:asym_preemptive}
Assume $\Lambda_v(\theta)$ is bounded, the asymptotic outage probability of source $i$ under single packet queue is
		\begin{align}
			\lim_{n \to \infty}\frac{1}{n} \log P_{out,i}^n(x) &\leq -I_v(x-b) \label{eqn:asym_upper_bound_preemptive}
		\end{align}
		where 
		\begin{equation}
            I_v(x) = \sup_{\theta>0} \left[ \theta x - \Lambda_v(\theta) \right].
        \end{equation}
\end{corollary}

\section{Discussion and Simulation Results}\label{sec:numerical and simulation}
In this section, we use computer simulation to verify our analysis. We then compare the two different queueing disciplines in different situations.

\begin{figure}
    \centering
    \includegraphics[width=0.32\textwidth]{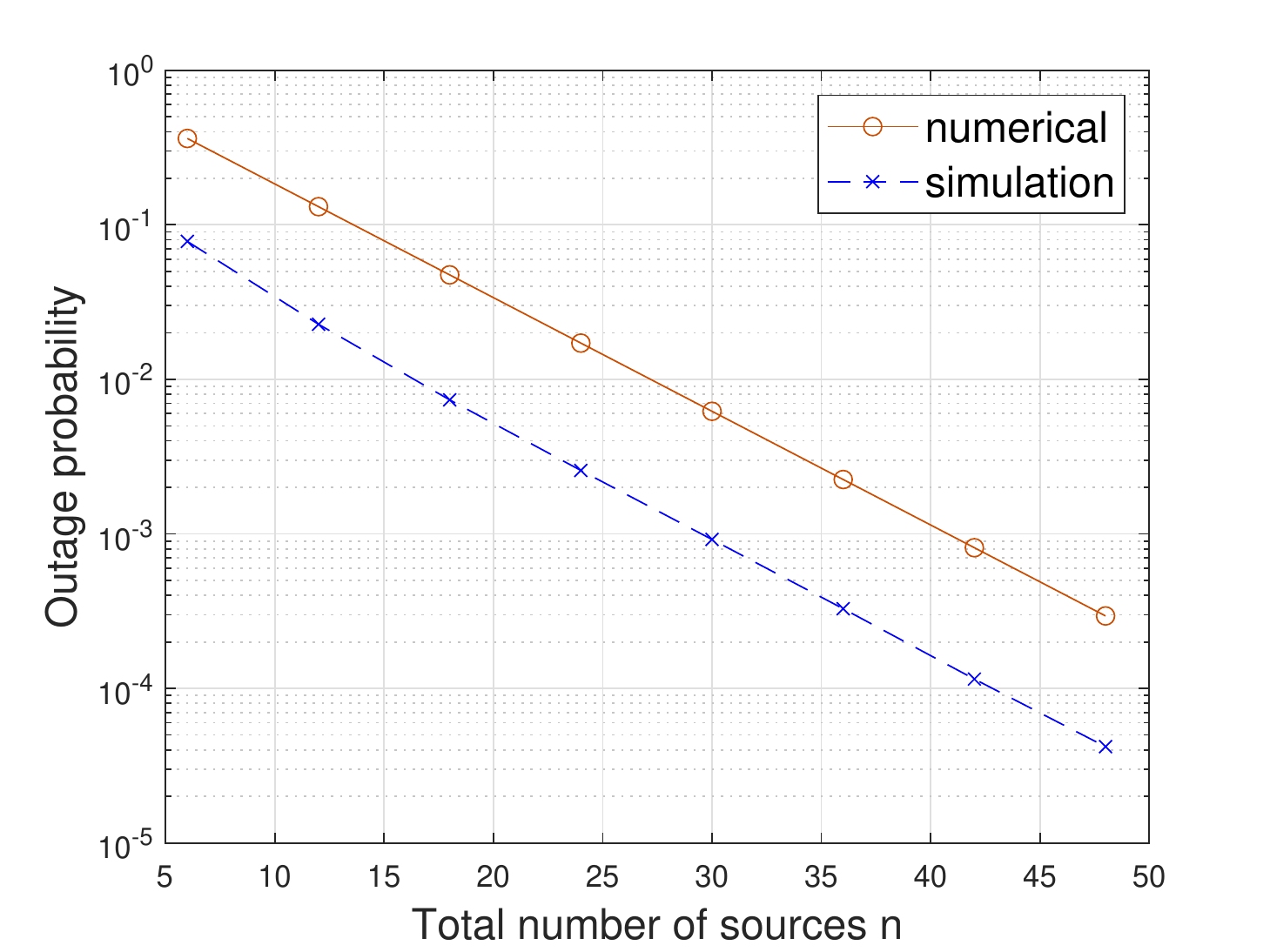}
    \caption{$P_{out}^n(x)$ against the $n$ in FCFS queue. We set $b=5$, $x=10$, and $V_i(k)\sim$ Poisson with $\lambda=3$.}
    \label{fig:simulation_FCFS}
    \vspace{-15pt}
\end{figure}

\begin{figure}
    \centering
    \includegraphics[width=0.32\textwidth]{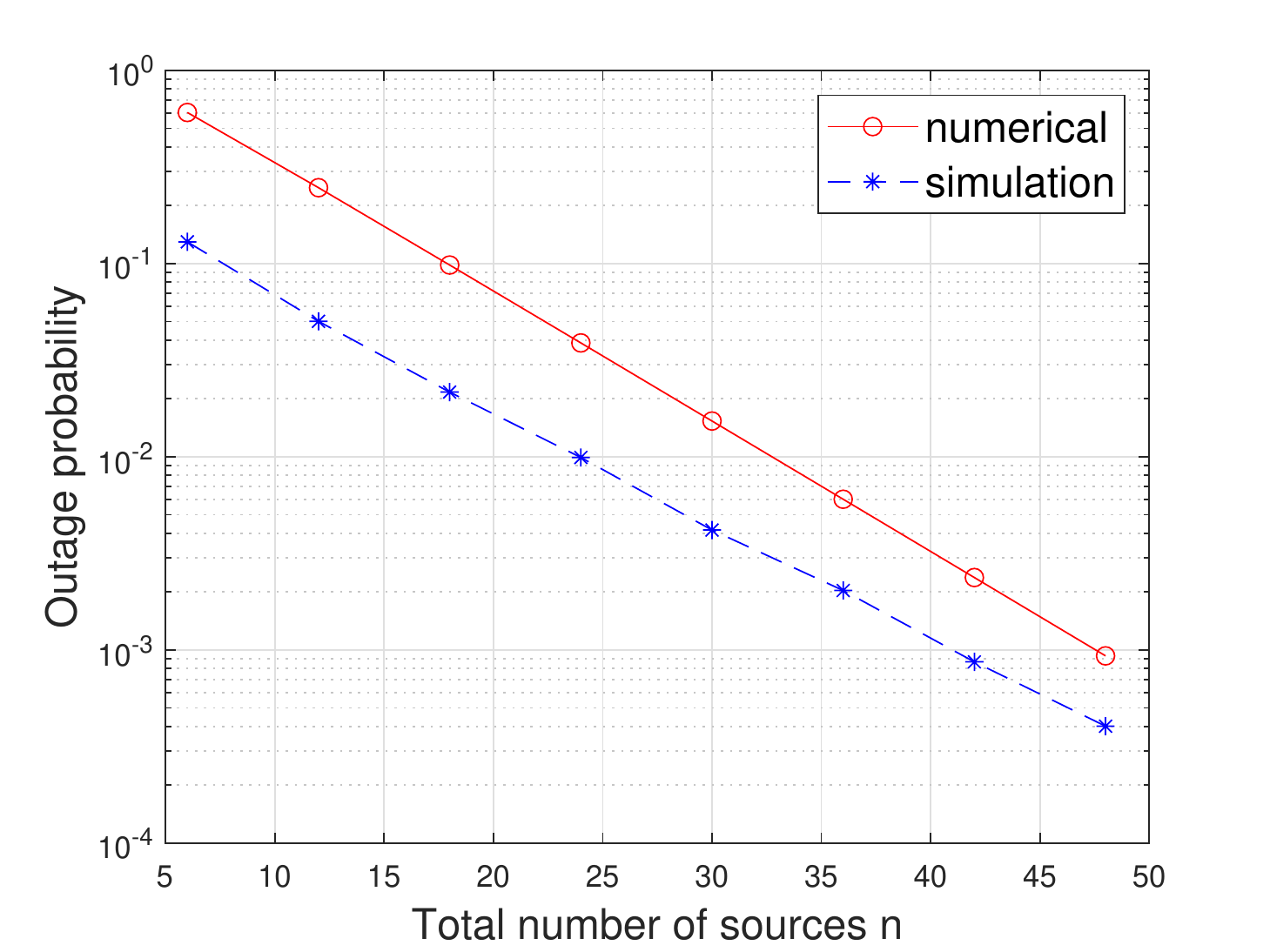}
    \caption{$P_{out}^n(x)$ against the $n$ in single packet queue. We set $b=5$, $x=10$, and $V_i(k)\sim$ Poisson with $\lambda=3$. }
    \label{fig:simulation_LCFS}
    \vspace{-15pt}
\end{figure}

In Figs.~\ref{fig:simulation_FCFS} and \ref{fig:simulation_LCFS}, we plot the outage probability against $n$ the number of sources in the system with FCFS queue and that with single packet queue, respectively. In both figures, the transmission time $V_i(k)$ is assumed to be i.i.d. according to the Poisson distribution with $\lambda=3$, the inter-arrival time is $nb$ with $b=5$, and the threshold is set to be $nx$ with $x=10$. Both the numerical results from our analysis and the simulation results from Monte-Carlo simulations are shown. We note that for the numerical results for the FCFS queue, we plot the asymptotic results in Corollary \ref{coro:asym_FCFS} because the precise constant $c$ is cumbersome to determine and does not really matter when $n$ is large. Moreover, for the numerical results for the single packet queue, we only plot the upper bound in Theorem~\ref{thm:upper_bound_preemptive} because a tight lower bound is lacking.

From Fig. \ref{fig:simulation_FCFS}, one observes that the simulation results and the numerical results display the same slope when $n$ is large. The gap between the two curves is due to the fact that the precise constant term is unknown. In Fig.~\ref{fig:simulation_LCFS}, it is again shown that the two curves have the same slope when $n$ is large. These results confirm the effectiveness and accuracy of our analysis.

Next, we compare the performance between two different queuing disciplines. We first focus on the case where the inter-arrival time is much larger than the total transmission time in a round. i.e. $nb > \sum_{u=1}^n V_u(k)$. Under this condition, the asymptotic decay rate in the FCFS queue can be expressed as \eqref{eqn:asym_FCFS} with $r=1$, which is given by
\begin{equation}\label{eqn:asym_prob FCFS under small transmission time}
    \lim_{n\to \infty} \frac{1}{n} P_{out,i}^n(x) = - \left[ \theta x - \theta b - \Lambda_v(\theta) \right].
\end{equation}
For the single packet queue case, by plugging \eqref{eqn:rate function preemptive} into \eqref{eqn:asym_upper_bound_preemptive}, we arrive at the same expression as \eqref{eqn:asym_prob FCFS under small transmission time}. This hints that the two queueing disciplines achieve the same asymptotic outage performance under this condition. On the other hand, for a general case, since the minimizer in \eqref{eqn:asym_FCFS} may not be $r=1$, our analysis concretizes the intuition that the single packet queue would result in a better outage performance than FCFS.


\section{Conclusion}\label{sec:conclusion}
In this work, we have considered a multi-source system that all sources wish to maintain the status of information. We have studied the performance guarantee under round robin scheduling by analyzing the outage probability of peak age exceeding a threshold. We have also provided the asymptotic decay rate in the multi-source regime to capture the behavior when the total number of sources becomes large. Simulation results have been provided to validate the analytic results. Our results have confirmed that the single packet queue outperforms the FCFS queue in terms of the outage probability. Moreover, the results have unveiled that the two disciplines lead to the same asymptotic decay rate when the inter-arrival time is much larger than the total transmission time.


\bibliographystyle{IEEEtran}
\bibliography{bib_5g.bib}

\onecolumn
\begin{appendices}

\section{Proof of Lemma~\ref{lemma:peak_age_formulation_FCFS}}\label{apx:proof_peak_age_formulation_FCFS}
\begin{IEEEproof}
	Substituting $W(k)$ in \eqref{eqn:A_i_FCFS}, with \eqref{eqn:W_i_FCFS} yields:
	\begin{align}\label{eqn:peak_age_first_iteration}
	A_i(k) &= \left[ W(k-1) + \sum_{u=1}^n V_u(k-1) -nb \right]^+ + \sum_{u=1}^i V_u(k) + nb \nonumber \\
	&= \max \left\{  W(k-1) + \sum_{u=1}^n V_u(k-1) + \sum_{u=1}^i V_u(k), \ \sum_{u=1}^i V_u(k) + nb \right\}
	\end{align}
	Substituting $W_(k-1)$ with \eqref{eqn:W_i_FCFS} again
	\begin{equation}
	\begin{aligned}
	A_i(k) &= \max \left\{ \left[ \vphantom{\sum_{u=1}^i V_u(k)} W(k-2) + \sum_{u=1}^n V_u(k-2) -nb \right]^+ + \sum_{u=1}^n V_u(k-1) + \sum_{u=1}^i V_u(k), \ \sum_{u=1}^i V_u(k) + nb \right\} \\
			&= \max \left\{ \left[ \vphantom{\sum_{u=1}^i V_u(k)} W(k-3) + \sum_{u=1}^n V_u(k-3) -nb \right]^+ + \sum_{r=k-2}^{k-1} \sum_{u=1}^n V_u(r) + \sum_{u=1}^i V_u(k) - nb, \right. \\
			&\left. \qquad\qquad \sum_{u=1}^n V_u(k-1) + \sum_{u=1}^i V_u(k), \ \sum_{u=1}^i V_u(k) + nb \right\} \\
			&= \max_{1\leq s\leq k} \left\{ \sum_{r=s}^{k-1} \sum_{u=1}^n V_u(r) + \sum_{u=1}^i V_u(k) - (k-s-1)nb \right\}.
	\end{aligned}
	\end{equation}
\end{IEEEproof}
The reason we traced all rounds before round $k$ is because we are looking for a way to represent the waiting time $W(k)$ in \ref{eqn:A_i_FCFS}.

\section{Proof of Theorem~\ref{thm:upper_bound_FCFS}}\label{apx:proof_upper_bound_FCFS}
\begin{IEEEproof}
	In this subsection, we derive the outage probability applying peak age in \eqref{eqn:peak_age_formulation_FCFS}. We can obtain:
	\begin{align*}
	Pr \left( \vphantom{\sum_{r=s}^{k-1} \sum_{u=1}^n V_u(r)} A_i(k) \geq nx \right) &= Pr \left( \max_{1\leq s\leq k} \left\{ \sum_{r=s}^{k-1} \sum_{u=1}^n V_u(r) + \sum_{u=1}^n V_u(k) - (k-s-1)nb \right\} \geq nx \right)\\
	&\overset{(a)}{\leq} \sum_{s=1}^k Pr \left( \vphantom{\sum_{r=s}^{k-1} \sum_{u=1}^n V_u(r)} \sum_{r=s}^{k-1} \sum_{u=1}^n V_u(r) + \sum_{u=1}^i V_u(k) - (k-s-1)nb \geq nx \right)\\
	&\overset{(b)}{\leq} \sum_{s=1}^k  \mathbb{E}\left[ \vphantom{\sum_{r=s}^{k-1} \sum_{u=1}^n V_u(r)} e^{\theta \left[ \sum_{r=s}^{k-1} \sum_{u=1}^n V_u(r) + \sum_{u=1}^i V_u(k) \right] } \right] \cdot e^{-(k-s-1)n\theta b} \cdot e^{-n\theta x}\\
	\end{align*}
	Inequation (a) uses the union bound, and inequation (b), uses the Chernoff bound with $\theta > 0$ being the parameter of the moment generating function. From the i.i.d assumption of  $V_i(k)$, we can have the following derivation:
	\begin{align*}
	&\sum_{s=1}^k  \mathbb{E}\left[ \vphantom{\sum_{r=s}^{k-1} \sum_{u=1}^n V_u(r)} e^{\theta \left[ \sum_{r=s}^{k-1} \sum_{u=1}^n V_u(r) + \sum_{u=1}^i V_u(k) \right] } \right] \cdot e^{-(k-s-1)n\theta b} \cdot e^{-n\theta x}\\
	&\overset{(c)}{=} \sum_{s=1}^k e^{ \left[ n\cdot (k-s)+i \right] \Lambda_v(\theta)} e^{-\theta (k-s-1)n\theta b} \cdot e^{-n\theta x}\\
	&\overset{(d)}{=} \sum_{r=1}^{k} e^{ n\cdot r \cdot \Lambda_v(\theta)} e^{-(n-i)\cdot \Lambda_v(\theta)} e^{- n \cdot (r-2) \theta b} \cdot e^{-n\theta x}\\
	&= \sum_{r=1}^{k} e^{-n\cdot r \cdot \left[ \frac{\theta x}{r} + \frac{r-2}{r} \theta b - \frac{r+(1-\alpha_i)}{r}\Lambda_v(\theta) \right]}\\
	\end{align*}
	where $\alpha_i = \frac{i}{n}$. In equality (c), we apply the log moment generating function $\Lambda_v(\theta) = \log \mathbb{E}[e^{\theta V_i(k)}$. In equation (d), we define $r=k-s+1$ and exchange the index of the summation. Next, in order to obtain a tight upper bound, we choose a specific $\theta^* = \sup_\theta \left\{ \frac{\theta x}{r} + \frac{r-2}{r} \theta b - \frac{r+(1-\alpha_i)}{r}\Lambda_v(\theta) \right\}$ to maximize the exponent. To simplify the expression, we define the rate function $I_v(x) = \sup_\theta \left\{ \theta x - \frac{r+(1-\alpha_i)}{r}\Lambda_v(\theta) \right\}$.
	
	Next, we let $k$ goes to infinite. We first separate the summation into two parts. Condition of $\theta b > \Lambda_v(\theta)$, one part of the summation becomes a geometric series. After some algebra, we can bound the outage probability $r_1$ times of maximum term.
	\begin{align}\label{eqn:outage prob two terms}
	\lim_{k\to \infty} Pr \left( \vphantom{\sum_{r=s}^{k-1} T(r)} A_n(k) \geq nx \right) 
	&\leq \sum_{r=1}^\infty e^{ -n \cdot r \cdot  I_v(\frac{x}{r}+\frac{r-2}{r}b) } \nonumber \\
	&\overset{(e)}{\leq} \sum_{r=1}^{r_1-1} e^{-n\cdot r \cdot I_v(\frac{x}{r}+\frac{r-2}{r}b)} + \sum_{r=r_1}^\infty e^{ -n \cdot r \cdot \left[ \theta b - \Lambda_v(\theta) \right] } \cdot e^{-n \left[ \theta x - 2\theta  b + (1-\alpha_i)\Lambda_v(\theta) \right]} \nonumber \\
	&= \sum_{r=1}^{r_1-1} e^{-n\cdot r \cdot I_v(\frac{x}{r}+\frac{r-2}{r}b)} + \frac{e^{-n \cdot r_1 \left[ \theta b - \Lambda_v(\theta) \right] }}{ 1-e^{-n \left[ \theta b - \Lambda_v(\theta) \right] } } \cdot e^{-n \left[ \theta x - 2\theta b + (1-\alpha_i)\Lambda_v(\theta) \right]} \nonumber \\
	&\overset{(f)}{\leq} \sum_{r=1}^{r_1-1} e^{-n\cdot r \cdot I_v(\frac{x}{r}+\frac{r-2}{r}b)} + e^{-n \cdot r_1 \left[ \theta b - \Lambda_v(\theta) \right] } \cdot e^{-n \left[ \theta x - 2\theta  b + (1-\alpha_i)\Lambda_v(\theta) \right]}
	\end{align}
	Since there exists a value for $r_1$ such that $r_1 > \arg \min_{r\geq 1} r \cdot I_v \left(\frac{x}{r}+\frac{r-2}{r}b \right)$. Inequation (e) holds by the definition of rate function $I_v(\cdot)$ below, for any $\theta$,
	\begin{align*}
	    I_v(\frac{x}{r}+\frac{r-2}{r}b) &\geq \theta \left( \frac{x}{r}+\frac{r-2}{r}b \right) - \frac{r+(1-\alpha_i)}{r}\Lambda_v(\theta)\\
	    &= \frac{1}{r} \left[ \vphantom{\frac{x}{r}} \theta x + (r-2)\theta b - r\cdot\Lambda_v(\theta) - (1-\alpha_i)\Lambda_v(\theta) \right]\\
	    &= \frac{1}{r}\left[ r (\vphantom{\frac{x}{r}}\theta b - \Lambda_v(\theta)) - 2\theta b -(1-\alpha_i)\Lambda_v(\theta) \right].
	\end{align*}
	and inequation (f), we remove the denominator of the second term as the upper bound. Next, we bound the exponent of \eqref{eqn:outage prob two terms} by $-n \cdot \min_{r\geq 1} r \cdot I_v(\frac{x}{r}+\frac{r-2}{r}b)$. We can have the following equation:
	\begin{align*}
	Pr \left( \vphantom{\sum_{r=s}^{k-1} T(r)} A_n(k) \geq nx \right)
	\leq r_1 \cdot e^{-n\cdot \min_{r\geq 1} \ r\cdot I_v(\frac{x}{r}+\frac{r-2}{r}b)}\\
	\end{align*}
	Finally, we complete the proof of the upper bound.
\end{IEEEproof}

\section{Proof of Theorem~\ref{thm:lower_bound_FCFS}}\label{apx:proof_lower_bound_FCFS}
In the lower bound part, the key idea is that we choose an arbitrary $s$ in \eqref{eqn:peak_age_formulation_FCFS}.
\begin{IEEEproof}
	\begin{align*}
	Pr \left( \vphantom{\sum_{r=s}^{k-1} \sum_{u=1}^n V_u(r)} A_n(k) \geq nx \right) &= Pr \left( \max_{1\leq s\leq k} \left\{ \sum_{r=s}^{k-1} \sum_{u=1}^n V_u(r) + \sum_{u=1}^i V_u(k) - (k-s-1)nb \right\} \geq nx \right)\\
	&\overset{(a)}{\geq} Pr \left( \vphantom{\sum_{r=s}^{k-1} \sum_{u=1}^n V_u(r)} \sum_{r=s}^{k-1} \sum_{u=1}^n V_u(r) + \sum_{u=1}^i V_u(k) - (k-s-1)nb \geq nx \right)\\
	&\overset{(b)}{\geq} \mathbb{E}\left[ \vphantom{\sum_{r=s}^{k-1} \sum_{u=1}^n V_u(r)} e^{\theta \left[ \sum_{r=s}^{k-1} \sum_{u=1}^n V_u(r) + \sum_{u=1}^i V_u(k) \right] } \right] \cdot e^{-(k-s-1)n\theta b} \cdot e^{-n\theta x} \cdot e^{-n \epsilon}\\
	&\overset{(c)}{=} \vphantom{\sum_{r=s}^{k-1} T(r)} e^{ n\cdot [r+(1-\alpha_i)] \cdot \Lambda_v(\theta)} e^{-n \cdot (r-2)\theta b} \cdot e^{-n\theta x} \cdot e^{-n \epsilon}\\
	&= \vphantom{\sum_{r=s}^{k-1} \sum_{u=1}^n V_u(r)} e^{-n\cdot r \cdot \left[ \frac{\theta x}{r} + \frac{r-2}{r} \theta b - \frac{r+(1-\alpha_i)}{r}\Lambda_v(\theta)+ \epsilon \right]}\\
	&\overset{(d)}{\geq} \vphantom{\sum_{r=s}^{k-1} \sum_{u=1}^n V_u(r)} e^{ -n \left[ r \cdot I_v(\frac{x}{r}+\frac{r-2}{r}b) + \epsilon \right] }
	\end{align*}
\end{IEEEproof}
In inequation (a), we choose an arbitrary $s$ as the lower bound. In inequation (b), we apply the Cramer-Chernoff theorem and $\epsilon>0$. Let $r=k-s+1$ in equation (c). In equation (d), we apply the rate function $I_v(\cdot)$ which we used in Appendix\ref{apx:proof_upper_bound_FCFS}, then we complete the proof of the lower bound.

\section{Proof of Theorem~\ref{thm:upper_bound_preemptive}}\label{apx:proof_upper_bound_preemptive}
\begin{IEEEproof}
In this section, we will derive the upper bound of the peak age under the preemptive queue. We first start from the expression of peak age in Lemma~\ref{lemma:peak_age_formulation_preemption}. Since the packet in the queue waiting to be transmitted will always be preempted when a new packet arrives, the waiting time in \eqref{eqn:peak_age_preemptive} will never be greater than the inter-arrival time $nb$. Based on this property, we get the upper bound of peak age below:
	\begin{equation}\label{eqn:peak_age_upper_bound_preemptive}
	\begin{aligned}
	A_i(k) &= W_i(k-1) + \sum_{u=i}^n V_u(k-1) + N_i(k-1) + \sum_{u=1}^i V_u(k)\\
			&\overset{(a)}{<} nb + \sum_{u=i}^n V_u(k-1) + N_i(k-1) + \sum_{u=1}^i V_u(k)\\
			&\overset{(b)}{=} nb + \sum_{u=i}^n V_u(k-1) + \sum_{u=1}^i V_u(k)
	\end{aligned}
	\end{equation}
In inequation (a), we apply $W_i(k-1) < nb$ and in equation (b), $N_i(k-1)$ is equal to zero since the waiting time is already greater than one inter-arrival $nb$, which ensures that we must have new packets to transmit.
After getting \eqref{eqn:peak_age_upper_bound_preemptive}, we start to derive the upper bound of the outage probability of peak age of source $i$ under preemptive queue.
	\begin{align*}
	Pr \left( A_n(k) \geq nx \right) &\leq Pr \left( nb + \sum_{u=i}^n V_u(k-1) + \sum_{u=1}^i V_u(k) \geq nx \right)\\
	&\overset{(c)}{\leq} \mathbb{E}\left[ \vphantom{\sum_{r=s}^{k-1} T(r)} e^{\theta \sum_{u=i}^n V_u(k-1) + \theta\sum_{u=1}^i V_u(k)} \right] \cdot e^{n\theta b} \cdot e^{-n\theta x}\\
	&\overset{(d)}{=} e^{(n+1)\Lambda_v(\theta)} \vphantom{\sum_{r=s}^{k-1} T(r)} \cdot e^{n\theta b} \cdot e^{-n\theta x}\\
	&= \vphantom{\sum_{r=s}^{k-1} T(r)} e^{-n\left[ \theta x - \theta b - \frac{n+1}{n}\Lambda_v(\theta)\right]}\\
    \end{align*}
In inequation (c), we apply the Chernoff bound and in equation (d) we assume the transmission time is i.i.d across users and packets and apply $\Lambda_v(\theta)$ which is defined in Appendix \ref{apx:proof_upper_bound_FCFS}. To get a tight upper bound, we apply the rate function $I_v^{(U)}(x) = \sup_\theta \left\{ \theta x - \frac{n+1}{n}\Lambda_v(\theta) \right\}$. Then we can achieve the equation below and we complete the proof of upper bound.
\begin{equation}
    Pr \left( A_i(k) \geq nx \right) \leq \vphantom{\sum_{r=s}^{k-1} T(r)}  e^{-n\cdot I_v(x-b)}
\end{equation}
\end{IEEEproof}

\end{appendices}

\end{document}